\documentclass[
 reprint,
 amsmath,amssymb,
 aps,
]{revtex4-2}

\usepackage{graphicx}
\usepackage{dcolumn}
\usepackage{bm}
\usepackage{hyperref}

\usepackage{float}
\usepackage{tikz}
\usepackage{tikz-feynman}
\tikzfeynmanset{compat=1.1.0}

\begin{document}

\preprint{APS/123-QED}

\title{Screening rho-meson mass in the presence of strong magnetic fields}

\author{L. A. Hern\'andez}
\author{Juan D. Mart\'inez-S\'anchez}%
\affiliation{%
Departamento de F\'isica, Universidad Aut\'onoma Metropolitana-Iztapalapa, Avenida San Rafael Atlixco 186, Ciudad de México 09340, Mexico.}%
\author{R. Zamora}
\affiliation{Instituto de Ciencias B\'asicas, Universidad Diego Portales, Casilla 298-V, Santiago, Chile.}
\affiliation{Facultad de Ingenier\'ia, Arquitectura y Dise\~no, Universidad San Sebasti\'an, Santiago, Chile.}


\begin{abstract}
We study the screening mass of the neutral rho-meson in the presence of strong magnetic fields using the Kroll-Lee-Zumino (KLZ) model. The rho-meson self-energy is computed at one-loop order within the lowest Landau level (LLL) approximation, considering the magnetic field as the dominant energy scale. Due to Lorentz symmetry breaking induced by the external field, we decompose the self-energy into three independent tensor structures, which give rise to three distinct modes. Additionally, the four-momentum splits into parallel and perpendicular components, leading to two types of screening masses: the parallel screening mass ( $p_0=0$ and $p_\perp \to 0$ ) and the perpendicular screening mass ( $p_0=0$ and $p_\parallel \to 0$ ). Our results show that the zero and perpendicular modes exhibit a monotonically increasing behavior with the magnetic field strength, whereas the parallel mode remains essentially constant. These findings provide new insights into the behavior of vector mesons in strongly magnetized media, with implications for QCD under extreme conditions.
\end{abstract}

\maketitle


\section{\label{sec1}Introduction}

Over the past few decades, the study of the QCD matter under extreme conditions has included analyses of strong magnetic fields and their effects on such systems. These investigations are motivated not only by theoretical curiosity but also by the fact that extremely strong magnetic fields are observed in various physical contexts. Examples include astronomical objects such as neutron stars~\cite{Ho:2011gy,Gusakov:2017uam,Igoshev:2021ewx}, the early universe~\cite{Grasso:2000wj,Subramanian:2009fu}, and experiments such as relativistic heavy-ion collisions~\cite{Voronyuk:2011jd,Bzdak:2011yy,McLerran:2013hla}. In these systems, magnetic field strengths on the order of the squared pion mass have been reported~\cite{Skokov:2009qp,Brandenburg:2021lnj,STAR:2023jdd}. 

Numerous studies have examined the influence of magnetic fields on the QCD phase transition~\cite{Andersen:2014xxa,Miransky:2015ava,Bandyopadhyay:2020zte,Hattori:2023egw,Ayala:2021nhx}, many of which were inspired by lattice QCD results showing that the pseudocritical temperature at zero baryon chemical potential decreases as the magnetic field strength increases. This phenomenon is now known as \textit{inverse magnetic catalysis}~\cite{Bali:2011qj,Bali:2012zg,Bali:2014kia}. Recently, various intriguing results have been reported in the literature; see~\cite{Adhikari:2024bfa} and the references therein. 

Since magnetic fields are present in the aforementioned physical systems, it is crucial to study their effects on the dynamics of QCD matter under extreme conditions. On one hand, this involves analyzing the modification of interaction strength due to magnetic fields alone or in combination with finite temperature and/or density~\cite{Ayala:2014uua,Ayala:2015bgv,Ayala:2016bbi,Ayala:2018wux,Fernandez:2024tuk}. On the other hand, the study of collective phenomena, which can be captured through the pole and screening masses of the degrees of freedom that constitute this kind of matter, remains an important research topic. These effects are considered in the presence of magnetic fields and/or within a thermal bath~\cite{Bali:2017ian, Ayala:2023llp}. 

The goal of this work is to focus on a particularly interesting property: the screening mass of the neutral rho-meson in the presence of magnetic fields, a quantity that has not yet been studied. However, several results exist regarding its pole mass. Lattice QCD results reported in~\cite{Bali:2017ian} describe the behavior of meson pole masses for both pions and rho-mesons in the presence of magnetic fields, for both neutral and charged states. For the neutral rho-meson, the pole mass increases as the magnetic field strength increases. In Ref.~\cite{Carlomagno:2022inu}, the authors studied the rho-meson pole mass as a function of the magnetic field using a two-flavor Nambu-Jona-Lasinio model. After a careful treatment, where pseudo-scalar-vector mixing played a key role, they found an increasing behavior of the pole mass with increasing magnetic field strength for all three spin projections. Similar results were obtained in Ref.~\cite{Liu:2014uwa}, where the authors employed the Nambu–Jona-Lasinio model, constructing mesons through an infinite sum of quark-loop chains using a random phase approximation. They calculated the polarization function at leading order in the $1/N_c$ expansion and found an increasing mass as a function of the magnetic field strength. In Ref.~\cite{Zhang:2016qrl}, using the $\Phi$-derivable approach in the Nambu–Jona-Lasinio model, the authors reported an increasing pole mass for the rho-meson as the magnetic field intensity increased. All these results are consistent with the qualitative behavior of the pole mass for the neutral rho-meson. Given this, we now turn our attention to the screening mass of the rho-meson. To our knowledge, there are no existing studies on this property when only magnetic fields are considered. In this work, we present the magnetic field dependence of the screening mass of the rho-meson, considering extremely strong magnetic fields within the Kroll-Lee-Zumino (KLZ) model. Due to the high magnetic field strengths considered, we employ the lowest Landau-level (LLL) approximation. 

This paper is organized as follows: in Section \ref{sec2}, we introduce the KLZ model. In Section \ref{sec3}, we compute all contributions to the rho-meson self-energy at one-loop order, incorporating the LLL approximation in propagators for charged fields in the loops. In Section \ref{sec4}, we write the equations for the screening mass and present their solutions. Finally, in Section \ref{sec5}, we discuss our results and provide conclusions.

\section{\label{sec2} Kroll-Lee-Zumino model}

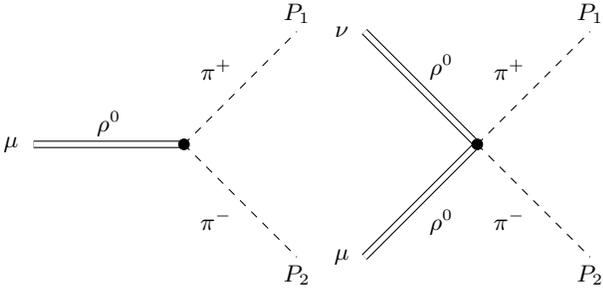
\begin{figure}[t]
    \centering
    \begin{tikzpicture}
        \node at (-2.3, 0) {$\mu$};    
        \draw [double distance=2pt] (-2,0) -- (0,0) node[midway, above] {$\rho^0$};
        
        \draw [dashed] (0,0) -- (1.5,1.5) node[midway, above left] {$\pi^+$} node[above] {$P_1$};
        
        \draw [dashed] (0,0) -- (1.5,-1.5) node[midway, below left] {$\pi^-$} node[below] {$P_2$};
        
        \filldraw (0,0) circle (2pt);
    \end{tikzpicture}
    \bigskip
    \begin{tikzpicture}
        \node at (-1.8, 1.5) {$\nu$};
        \node at (-1.8, -1.5) {$\mu$};
        
        \draw [double distance=2pt] (-1.5,1.5) -- (0,0) node[midway, above right] {$\rho^0$};
        \draw [double distance=2pt] (-1.5,-1.5) -- (0,0) node[midway, below right] {$\rho^0$};
        
        \draw [dashed] (0,0) -- (1.5,1.5) node[midway, above left] {$\pi^+$} node[above] {$P_1$};
        \draw [dashed] (0,0) -- (1.5,-1.5) node[midway, below left] {$\pi^-$} node[below] {$P_2$};
        
        \filldraw (0,0) circle (2pt);
    \end{tikzpicture}    
    \caption{Feynman diagrams of the vertex in the KLZ model. The one of the left is the interaction $\rho^0 \rightarrow \pi^+ + \pi^-$, which vertex is $ig(P_1+P_2)^\mu$, an the one on the right is the interaction $\rho^0_\nu + \rho^0_\mu \rightarrow \pi^+ + \pi^-$, which vertex is $2i g^2 g^{\mu\nu}$.}
    \label{fig1}
\end{figure}

The KLZ model is a quantum field theory of strong interactions, whose degrees of freedom are pions and a massive neutral rho-meson~\cite{Kroll:1967it}. Although the rho-meson is a massive gauge boson, this model remains renormalizable due to the coupling of this gauge boson to a conserved current~\cite{vanHees:2003dk}. Moreover, this version of the model is an Abelian theory. One of its most important features is that it provides a justification, from a quantum field theory perspective, for the Vector Meson Dominance model~\cite{Sakurai:1960ju}. The Lagrangian density of the KLZ model is given by
\begin{align}
    \mathcal{L}&=\partial_\mu \pi^- \partial^\mu \pi^+-m_\pi^2\pi^-\pi^+-\frac{1}{4}\rho_{\mu\nu}\rho^{\mu\nu}+\frac{1}{2}M_\rho^2\rho_\mu\rho^\mu \nonumber \\
    &+g (i\pi^-\overleftrightarrow{\partial_\mu}\pi^+)\rho^\mu+g^2\rho_\mu\rho^\mu\pi^-\pi^+,
    \label{LagrangianKLZ}
\end{align}
where $m_\pi$ and $M_\rho$ are the masses of the pions and the rho-meson, respectively. The vector field $\rho^\mu$ corresponds to the neutral rho-meson, while the charged pions are represented by $\pi^\pm$. The field strength tensor is given by $\rho_{\mu\nu}=\partial_\mu\rho_\nu-\partial_\nu\rho_\mu$, and $g$ is the coupling constant between the charged pions and the neutral rho-meson. For this model, we have two different interaction vertices, which are depicted in Fig.~\ref{fig1}. Finally, an external uniform and constant magnetic field can be included in the model, introducing a covariant derivative in the Lagrangian density, namely
\begin{equation}
    \partial_\mu \rightarrow D_\mu=\partial_\mu+iqA_\mu,
    \label{covariantder}
\end{equation}
where $A_\mu$ is the vector potential corresponding to an external magnetic field along the $\hat{z}$ axis, and $q$ is the charge of the field.

Using the Feynman rules of the KLZ model, we proceed in the next section to compute the one-loop contributions to the neutral rho-meson self-energy.

\section{\label{sec3} Self-energy of the rho-meson}
\begin{figure}[t]
    \centering
    \begin{tikzpicture}
        \draw[double distance=2pt] (-2.8, 0) -- (2.8, 0) node[midway, below] {$\rho^0$}; 
        \node at (-3, 0) {$\mu$};
        \node at (3, 0) {$\nu$};
        \node at (-2, 0.4) {$\vec{p}$};
        \node at (2, 0.4) {$\vec{p}$};
        
        \draw[dashed] (0, 1.06) circle (1);

        \node at (0, 2.36) {$\pi^\pm$};
        \node at (0, 1.7) {$\vec{k}$};
        
        \filldraw (0, 0) circle (2pt);
    \end{tikzpicture}
    \caption{Tadpole diagram for the one-loop correction to the rho-meson self-energy.}
    \label{tadpolefigure}
\end{figure}
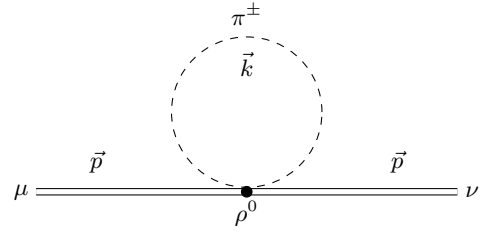
The neutral rho-meson self-energy at one-loop has two contributions under the influence of the magnetic fields: the tadpole diagram, depicted in Fig.~\ref{tadpolefigure}, and the
$\pi^+\pi^-$ diagram, depicted in Fig.~\ref{pionsfigure}. Both contributions involve charged particles in the loop. The first contribution is expressed as 
\begin{figure}[b]
    \centering
    \begin{tikzpicture}
        \draw[double distance=2pt] (-2.8, 0) -- (-1, 0) node[midway, below] {$\rho^0$}; 
        \node at (-3, 0) {$\mu$};
        \node at (-2, 0.4) {$\vec{p}$};

        \draw[double distance=2pt] (1, 0) -- (2.8, 0) node[midway, below] {$\rho^0$}; 
        \node at (3, 0) {$\nu$};
        \node at (2, 0.4) {$\vec{p}$};
        
        \draw[dashed] (0, 0) circle (1);

        \node at (0, 1.3) {$\pi^+$};
        \node at (0, 0.7) {$\vec{k}$};

        \node at (0, -1.3) {$\pi^-$};
        \node at (0, -0.65) {$\vec{p}-\vec{k}$};        
        
        \filldraw (-1, 0) circle (2pt);
        \filldraw (1, 0) circle (2pt);
    \end{tikzpicture}
    \caption{$\pi^+ \pi^-$ diagram for the one-loop correction to the rho-meson self-energy.}
    \label{pionsfigure}
\end{figure}
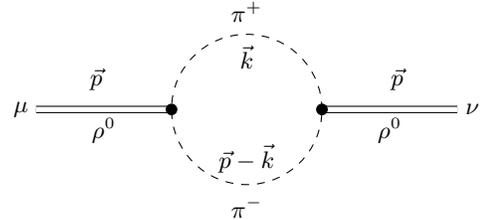
\begin{equation}
    i\Sigma^\text{t}_{\mu\nu}=2ig^2g^{\mu\nu}\int \frac{d^4k}{(2\pi)^4}D^{\text{LLL}}(k),
    \label{Tadpole_self-energy}
\end{equation}
while for the second contribution, we have
\begin{align}
   i\Sigma^\pm_{\mu\nu}=\int \frac{d^4k}{(2\pi)^4} &ig(2k-p)_\mu D^{\text{LLL}}(p-k)\nonumber \\
   &\times ig(2k-p)_\nu D^{\text{LLL}}(k).
   \label{pions_self-energy}
\end{align}
The propagator for a charged scalar field in the presence of a homogeneous and constant magnetic field in the lowest Landau level approximation (the strength of this field is the largest energy scale) is
\begin{equation}
  D^{LLL}(k)=\frac{2i e^{-\frac{k_{\perp}^2}{2 eB}}}{k_{\parallel}^2-m_B^2+i\epsilon},  
  \label{scalar_prop_LLL}
\end{equation}
where $m_B^2=m_\pi^2+eB$ is a magnetic mass. Equations~(\ref{Tadpole_self-energy}) and~(\ref{pions_self-energy}) exhibit a tensor structure in the Lorentz space, which is appropriate given the vector nature of the rho meson.

Since the magnetic field is aligned along the $\hat{z}$-axis, the four-momentum space is split into parallel $(x^0,x^3)$ and perpendicular $(x^1,x^2)$ components. The metric tensor is decomposed as
\begin{equation}
    g^{\mu\nu}=g^{\mu\nu}_\parallel+g^{\mu\nu}_\perp,
\end{equation}
and the squared four-vector follows
\begin{equation}
    X^\mu X_\mu=X^2=X_\parallel^2-X_\perp^2,
\end{equation}
where $X_\parallel^2=X_0^2-X_3^2$ and $X_\perp^2=X_1^2+X_2^2$. The presence of a uniform magnetic field breaks Lorentz symmetry, leading to the formation of four linearly independent tensor structures. There is a freedom to choose the basis on which we want to write the self-energy of the rho-meson. For this work, we notice that there are four possible independent tensors $p^\mu p^\nu, \ b^\mu b^\nu, \ p^\mu b^\nu + p^\nu b^\mu$ and $g^{\mu\nu}$, where $b^\mu$ is the four-vector for the direction of the magnetic field. However, the rho-meson is a gauge vector and the Ward-Takahashi identity must be satisfied
\begin{equation}
    p_\mu p_\nu \Sigma^{\mu\nu}=0,
    \label{transversality}
\end{equation}
and the self-energy must be transverse. Therefore, a convenient choice of basis to express the rho-meson self-energy needs only three linearly independent tensor structures, given by
\begin{equation}
    \Sigma^{\mu\nu}=P_\parallel \Sigma_\parallel^{\mu\nu}+P_\perp \Sigma_\perp^{\mu\nu}+P_0 \Sigma_0^{\mu \nu},
    \label{tensorbasis}
\end{equation}
where
\begin{align}
    \Sigma_\parallel^{\mu \nu}&=g_\parallel^{\mu \nu}-\frac{p_\parallel^\mu p_\parallel^\nu}{p^2_\parallel}, \nonumber \\
    \Sigma_\perp^{\mu \nu}&=g_\perp^{\mu \nu}+\frac{p_\perp^\mu p_\perp^\nu}{p^2_\perp}, \nonumber \\
    \Sigma_0^{\mu \nu}&=g^{\mu \nu}-\frac{p^\mu p^\nu}{p^2}-\Pi_\parallel^{\mu \nu}-\Pi_\perp^{\mu \nu}.
    \label{elementsbasis}
\end{align}
This orthonormal basis ensures gauge invariance~\cite{Hattori:2017xoo}, and is inspired by studies conducted for the polarization tensor either in QED or QCD~\cite{Hattori:2017xoo,Ayala:2019akk,Ayala:2020wzl}.

Once we have defined the elements of the tensor basis, we proceed to compute the coefficients for each of the basis elements.

\subsection{Coefficient $P_0$}

The first coefficient to be computed is the one corresponding to the tensor $\Sigma_0^{\mu \nu}$. It is obtained from
\begin{align}
 P_0&=\Sigma_0^{\mu \nu} (i\Sigma^\text{t}_{\mu\nu}+i\Sigma^\pm_{\mu\nu}) \nonumber \\
 &=P_0^\text{t}+P_0^\pm,
 \label{P0definition}
\end{align}

From Eq.~(\ref{P0definition}), we start the calculation of the contribution $P_0^\text{t}$. After the contraction between $\Sigma_0^{\mu \nu}$ and $i\Sigma^t_{\mu\nu}$, we obtain the coefficient
\begin{equation}
    P_0^\text{t}= 2ig^2\int \frac{d^4k}{(2\pi)^4}D^{\text{LLL}}(k),
    \label{coeff_P0}
\end{equation}
substituting the propagator of Eq.~(\ref{scalar_prop_LLL}) into Eq.~(\ref{coeff_P0}), we get
\begin{equation}
    P_0^\text{t}=-4g^2 \int \frac{d^2k_\parallel}{(2\pi)^2} \int \frac{d^2k_\perp}{(2\pi)^2} \frac{ e^{-\frac{k_{\perp}^2}{2 eB}}}{k_{\parallel}^2-m_B^2+i\epsilon}.
\end{equation}
We have split the integrals into two terms, parallel and perpendicular components. We integrate over the perpendicular components and obtain
\begin{equation}
    P_0^\text{t}=-g^2 \frac{eB}{\pi}\int \frac{d^2k_\parallel}{(2\pi)^2} \frac{1}{k_{\parallel}^2-m_B^2+i\epsilon}.
\end{equation}
The next step corresponds to integrate over the parallel components; however, it has a UV divergence. In order to perform this integral, we use dimensional regularization and $\overline{MS}$ scheme, and we have
\begin{eqnarray}
    P_0^\text{t}= i\frac{g^2 eB}{4 \pi^2} \ln\left(\frac{\mu^2}{m_B^2}\right),
\end{eqnarray}
where $\mu$ is the renormalization scale. Notice that this term is not independent of $eB$; it arises as a consequence of the lowest Landau level approximation, where the vacuum cannot be isolated. This phenomenon occurs in every UV-divergent term that appears in this work.

We now compute $P_0^\pm$, the coefficient after the inner product between $\Sigma_0^{\mu \nu}$ and $i\Sigma_{\mu \nu}^\pm$ tensors is the following
\begin{eqnarray}
P_0^\pm&=&- 4 g^2 \int \frac{d^4k}{(2\pi)^4} \left( \frac{p^2k^2}{p^2} - \frac{(p \cdot k)^2  }{p^2}\right) \nonumber \\ 
&\times& D^{LLL}(p-k)D^{LLL}(k).
\label{P0_initial}
\end{eqnarray}
We substitute the expression of the propagator from Eq.~(\ref{scalar_prop_LLL}) into Eq.~(\ref{P0_initial}), express the four momentum in terms of $k_\parallel$, $k_1$ and $k_2$, and obtain
\begin{align}
P_0^\pm&= \frac{16 g^2}{p^2} \int \frac{d^2k_{\parallel}dk_1dk_2}{(2\pi)^4} \Bigl(p^2k_{\parallel}^2-p^2(k_1^2+k_2^2)\nonumber \\
&-(p_{\parallel} \cdot k_{\parallel})^2+2p_{\parallel}\cdot k_{\parallel} (p_1k_1+p_2k_2) - p_1^2 k_1^2  \nonumber \\
&- 2p_1k_1p_2k_2 -p_2^2k_2^2\Bigr) \frac{e^{-(k_{\perp}^2+(p-k)_{\perp}^2)/(2eB)}}{(k_{\parallel}^2-m_B^2)((p-k)_{\parallel}^2-m_B^2)}. \nonumber \\
\end{align}
We perform the integrals over $k_1$ and $k_2$. Hence, $P_0^\pm$ becomes
\begin{align}
P_0^\pm&= -\frac{g^2 eB }{2p^2\pi} e^{-\frac{p_{\perp}^2}{2 eB}} \int \frac{d^2k_{\parallel}}{(2\pi)^2}  \Bigl(  p_{\perp}^2 (p^2+p_{\perp}^2) -4 k_{\parallel}^2 p^2 \nonumber \\ 
&+eB(2p^2 + p_{\perp}^2 -4 p_{\perp}^2 (p_{\parallel} \cdot k_{\parallel})+4(p_{\parallel} \cdot k_{\parallel})^2)\Bigr) \nonumber \\
&\times \frac{1}{(k_{\parallel}^2-m_B^2)((p-k)_{\parallel}^2-m_B^2)}.
\end{align}
In order to calculate the integral over $k_{\parallel}$, we use Feynman parametrization, we make the change of variable $k_{\parallel}=l_{\parallel}-(1-x)p_{\parallel}$ and get
\begin{align}
P_0^\pm&= -\frac{g^2 eB }{2p^2\pi} e^{-\frac{p_{\perp}^2}{2 eB}} \int_0^1dx\int \frac{d^2l_{\parallel}}{(2\pi)^2}  \Bigl(  p_{\perp}^2 p_{\parallel}^2 \nonumber\\
&+ eB(p_{\parallel}^2 + p^2) -4 p_{\perp}^2 p_{\parallel} \cdot (l_{\parallel}+(1-x)p_{\parallel})\nonumber \\
&+4(p_{\parallel} \cdot (l_{\parallel}+(1-x)p_{\parallel}))^2 -4 (l_{\parallel}+(1-x) p_{\parallel})^2 p^2 \Bigr) \nonumber \\
&\times \frac{1}{(l_{\parallel}^2-\Delta)^2},
\end{align}
with $\Delta=m_B^2-x(1-x)p_{\parallel}^2$. Now, we proceed to perform the integral over the parallel components; however, we first make a Wick rotation, $l_0 \rightarrow il_4$, and integrate over $l_3$
\begin{eqnarray}
P_0^\pm&=& -\frac{ig^2 eB}{8\pi p^2} e^{-\frac{p_{\perp}^2}{2 eB}} \int_0^1dx\int \frac{dl_4}{(2\pi)} \Biggl( \frac{4 (p^2+p_3^2)}{(l_4^2+\Delta)^{1/2}} \nonumber \\
&+& \frac{4 l_4^2 (p^2-p_0^2)}{(l_4^2+\Delta)^{3/2}} +\Bigl[p_{\perp}^2 p_{\parallel}^2 -4 (1-x)^2 p_{\parallel}^2p^2  \nonumber \\
&+& eB(p_{\parallel}^2+p^2)-4p_{\perp}^2p_{\parallel}^2(1-x)+ 4 (1-x)^2 p_{\parallel}^4 \Bigr]\Biggr) \nonumber \\
&\times&\frac{1}{(l_4^2+\Delta)^{3/2}}.
\label{P_0_pm_wick_rotation}
\end{eqnarray}
Note that in Eq.~(\ref{P_0_pm_wick_rotation}) we have three terms, the first two have a UV divergence, which we handle with dimensional regularization and use $\overline{MS}$ scheme. After completing the integral over $l_4$, it becomes
\begin{eqnarray}
P_0^\pm&=& -\frac{ig^2 eB}{8\pi p^2} e^{-\frac{p_{\perp}^2}{2 eB}} \int_0^1dx \nonumber \\
&\times&\Biggl( 2 (p^2-p_{\perp}^2) \ln\left(\frac{\mu^2}{m_B^2-x(1-x)p_{\parallel}^2}\right) \nonumber \\
&-&4(p^2-p_0^2) + \frac{(p_{\parallel}^2p_{\perp}^2 (1-2x)^2+eB(p_{\parallel}^2+p_{\perp}^2))}{m_B^2-x(1-x)p_{\parallel}^2} \Biggr). \nonumber \\
\end{eqnarray} 
The last step to perform is the integral over the Feynman parameter $x$, once it is calculated, we obtain
\begin{eqnarray}
P_0^\pm&=&-\frac{i g^2 (eB) e^{-\frac{p_{\perp}^2}{2 eB}} }{2 \pi^2 p^2}\Biggl[ p_0^2-p^2 -\frac{p_{\perp}^2}{2} \ln\left(\frac{\mu^2}{m_B^2}\right) \nonumber \\
&&+\frac{eB p_{\parallel}}{\sqrt{4 m_B^2-p_{\parallel}^2}} \arctan\left(\frac{ p_{\parallel}}{\sqrt{4 m_B^2-p_{\parallel}^2}}\right)  \Biggr]. \nonumber \\
\end{eqnarray}
Since we have at hand $P_0^{\text{t}}$ and $P_0^\pm$, we are able to write
\begin{eqnarray}
 P_0&=&P_0^\text{t}+P_0^\pm \nonumber \\
&=&-\frac{i g^2 (eB) e^{-\frac{p_{\perp}^2}{2 eB}} }{2 \pi^2 p^2}\Biggl[ p_0^2-p^2 -\frac{p_{\perp}^2}{2} \ln\left(\frac{\mu^2}{m_B^2}\right) \nonumber \\
&&+\frac{eB p_{\parallel}}{\sqrt{4 m_B^2-p_{\parallel}^2}} \arctan\left(\frac{ p_{\parallel}}{\sqrt{4 m_B^2-p_{\parallel}^2}}\right) \nonumber \\
&&-\frac{p^2}{2} \ln\left(\frac{\mu^2}{m_B^2}\right) e^{\frac{p_{\perp}^2}{2 eB}} \Biggr].
\label{finalP0}
\end{eqnarray}

\subsection{Coefficient $P_{\perp}$}

The next coefficient to be computed is $P_{\perp}$ which comes from the contraction of the basis element $\Pi_\perp^{\mu\nu}$ with $i\Sigma_{\mu\nu}^t$ and $i\Sigma_{\mu\nu}^\pm$. Then, we have
\begin{align}
 P_{\perp}&=\Pi_{\perp}^{\mu \nu} (i\Sigma^\text{t}_{\mu\nu}+i\Sigma^\pm_{\mu\nu}) \nonumber \\
 &=P_{\perp}^\text{t}+P_{\perp}^\pm.
 \label{contraction_perp}
\end{align}
The way to compute the coefficient $P_{\perp}$ is completely analogous to the one that we show for $P_0$. We start with $P_\perp^\text{t}$, it is written as
\begin{eqnarray}
    P_{\perp}^\text{t}= 2ig^2\int \frac{d^4k}{(2\pi)^4}D^{\text{LLL}}(k).
    \label{coeffP_perp}
\end{eqnarray}
We notice that the Eq.~(\ref{coeffP_perp}) is equal to Eq.~(\ref{coeff_P0}), hence we obtain 
\begin{eqnarray}
    P_{\perp}^\text{t}= i\frac{g^2 eB}{4 \pi^2} \ln\left(\frac{\mu^2}{m_B^2}\right),
    \label{finalptperp}
\end{eqnarray}
where we handle in the same manner the integration over the four momenta and the respective UV divergence as we did for $P_0^\text{t}$.

Now, we proceed to calculate the contribution $P_{\perp}^\pm$. To accomplish this, we first perform the contraction between $\Pi_\perp^{\mu\nu}$ and $i\Sigma_{\mu\nu}^\pm$, and we get
\begin{eqnarray}
 P_{\perp}^\pm&=&-\frac{16g^2}{p_{\perp}^2} \int \frac{d^4k}{(2\pi)^4}  (k_2 p_1- k_1 p_2)^2\nonumber \\
 &\times& \frac{e^{-(k_{\perp}^2+(p-k)_{\perp}^2)/(2eB)}}{(k_\parallel^2-m_B^2)((p-k)_\parallel^2-m_B^2)},
\end{eqnarray}
as a first step, we integrate over the two components $d^2k_{\perp}$, where we obtain
\begin{eqnarray}
 P_{\perp}^\pm&=&-\frac{g^2(eB)^2}{2\pi} e^{-\frac{p_{\perp}^2}{2eB}} \nonumber \\
 &\times&\int \frac{d^2k_{\parallel}}{(2 \pi)^2}  \frac{1}{(k_\parallel^2-m_B^2)((p-k)_\parallel^2-m_B^2)}.
\end{eqnarray}
For the parallel components, we use Feynman parametrization and implement the following change of variable $k_{\parallel}^2=l_{\parallel}^2- p_{\parallel}(1-x)$, then the expression becomes 
\begin{eqnarray}
 P_{\perp}^\pm&=&-\frac{g^2(eB)^2}{2\pi} e^{-\frac{p_{\perp}^2}{2eB}}\int_0^1dx\int \frac{d^2l_{\parallel}}{(2 \pi)^2}  \frac{1}{(l_\parallel^2-\Delta)^2},
\end{eqnarray}
we now integrate $d^2l_{\parallel}$ and obtain
\begin{eqnarray}
 P_{\perp}^\pm&=&-i\frac{g^2(eB)^2}{8\pi^2} e^{-\frac{p_{\perp}^2}{2eB}}\int_0^1dx  \frac{1}{m_B^2-x(1-x)p_{\parallel}^2}.
\end{eqnarray}
Finally, we perform the integral over the Feynman parameter $x$, and the coefficient is
\begin{align}
 P_{\perp}^\pm&=-i\frac{ g^2 (eB)^2 }{2 \pi^2}
 \frac{e^{-\frac{p_{\perp}^2}{2 eB}}}{p_{\parallel}\sqrt{4 m_B^2-p_{\parallel}^2}}\nonumber \\
&\times \arctan\left(\frac{ p_{\parallel}}{\sqrt{4 m_B^2-p_{\parallel}^2}}\right) .
\label{finalpperppm}
\end{align}

With the Eqs.~(\ref{finalptperp}) and~(\ref{finalpperppm}) at hand, we are able to write the coefficient corresponding to the basis element $\Pi^{\mu\nu}_\perp$ 
\begin{eqnarray}
&& P_{\perp}=-\frac{i g^2 (eB) e^{-\frac{p_{\perp}^2}{2 eB}} }{2 \pi^2}\Biggl[ -\frac{1}{2} \ln\left(\frac{\mu^2}{m_B^2}\right)e^{\frac{p_{\perp}^2}{2 eB}}
\nonumber \\
&&+\frac{eB}{p_{\parallel}\sqrt{4 m_B^2-p_{\parallel}^2}} \arctan\left(\frac{ p_{\parallel}}{\sqrt{4 m_B^2-p_{\parallel}^2}}\right) \Biggr].
\label{finalPperp}
\end{eqnarray}

\subsection{Coefficient $P_{\parallel}$}

The last coefficient to be calculated is the corresponding to the basis element $\Pi^{\mu\nu}_\parallel$. This is
\begin{align}
 P_{\parallel}&=\Pi_{\parallel}^{\mu \nu} (i\Sigma^\text{t}_{\mu\nu}+i\Sigma^\pm_{\mu\nu}) \nonumber \\
 &=P_{\parallel}^\text{t}+P_{\parallel}^\pm,
\end{align}
where $\Pi^{\mu\nu}_\parallel$, $i\Sigma^\text{t}_{\mu\nu}$ and $i\Sigma^\pm_{\mu\nu}$ are given by Eqs.~(\ref{elementsbasis}), (\ref{Tadpole_self-energy}) and~(\ref{pions_self-energy}), respectively. 

The path we follow to carry out the calculation of this last coefficient is in the same fashion as we did in the two previous cases. For the tadpole contribution, after performing the contraction between $\Pi^{\mu \nu}_\parallel$ and $i\Sigma_{\mu\nu}^t$, we obtain
\begin{equation}
    P^{t}_\parallel=2ig^2\int \frac{d^4k}{(2\pi)^4}D^{\text{LLL}}(k).
    \label{coefPtparallel}
\end{equation}
We know the result of the integral over the four momenta in Eq.~(\ref{coefPtparallel}), it is exactly the same as $P^t_0$ and $P^t_\perp$. Therefore, this contribution is
\begin{eqnarray}
    P_{\parallel}^\text{t}= i\frac{g^2 (eB)}{4 \pi^2} \ln\left(\frac{\mu^2}{m_B^2}\right).
    \label{Pparallelt}
\end{eqnarray}
For the second contribution, $P_{\parallel}^\pm$, we perform the contraction between $\Pi^{\mu \nu}_\parallel$ and $i\Sigma_{\mu\nu}^\pm$ and get
\begin{eqnarray}
 P_{\parallel}^\pm&=&-\frac{16g^2}{p_{\parallel}^2} \int \frac{d^4k}{(2\pi)^4}  (k_3 p_0- k_0 p_3)^2\nonumber \\
 &\times& \frac{e^{-(k_{\perp}^2+(p-k)_{\perp}^2)/(2eB)}}{(k_\parallel^2-m_B^2)((p-k)_\parallel^2-m_B^2)},
\end{eqnarray}
where we first integrate over $k_{\perp}$, then we implement the Feynman parametrization for the parallel components, and so we arrive at the expression
\begin{eqnarray}
 P_{\parallel}^\pm&=&-\frac{g^2(eB)}{2\pi p_{\parallel}^2} e^{-\frac{p_{\perp}^2}{2eB}}\int_0^1dx\int \frac{d^2l_{\parallel}}{(2 \pi)^2}  \frac{l_3^2p_0^2+l_0^2p_3^2}{(l_\parallel^2-\Delta)^2}. \nonumber \\
\end{eqnarray}
Then, performing the integral over $d^2l_{\parallel}$ and applying dimensional regularization and $\overline{MS}$ scheme, we obtain
\begin{eqnarray}
&& P_{\parallel}^\pm=-\frac{i g^2 (eB) e^{-\frac{p_{\perp}^2}{2 eB}} }{2 \pi^2}\Biggl[ \frac{1}{2} \ln\left(\frac{\mu^2}{m_B^2}\right) \nonumber \\
&&-\frac{\sqrt{4 m_B^2-p_{\parallel}^2}}{p_{\parallel}}\arctan\left(\frac{ p_{\parallel}}{\sqrt{4 m_B^2-p_{\parallel}^2}}\right) +1 \Biggr], \nonumber \\
\label{Pparallel+-}
\end{eqnarray}
Therefore, combining both contributions, from Eqs.~(\ref{Pparallelt}) and~(\ref{Pparallel+-}), we finally get
\begin{align}
 P_{\parallel}&=-\frac{i g^2 (eB) e^{-\frac{p_{\perp}^2}{2 eB}} }{2 \pi^2}\Biggl[ 1 -\frac{1}{2} \ln\left(\frac{\mu^2}{m_B^2}\right)e^{\frac{p_{\perp}^2}{2 eB}}\nonumber \\
&-\frac{\sqrt{4 m_B^2-p_{\parallel}^2}}{p_{\parallel}}\arctan\left(\frac{ p_{\parallel}}{\sqrt{4 m_B^2-p_{\parallel}^2}}\right) +\frac{1}{2} \ln\left(\frac{\mu^2}{m_B^2}\right) \Biggr]. \nonumber \\
\label{finalPparallel}
\end{align}

\section{\label{sec4} Screening mass}

In Sec.~\ref{sec3}, we present a detailed calculation of each coefficient accompanying the basis elements used to express the self-energy of the neutral rho-meson at one-loop order. With these results, we are now able to compute the screening masses for each of the three components of the self-energy. We begin by writing the propagator of the neutral rho-meson in the chosen basis
\begin{equation}
    D_\rho^{\mu\nu}=D_\parallel \Pi_\parallel^{\mu\nu}+D_\perp \Pi_\perp^{\mu\nu}+D_0 \Pi_0^{\mu\nu},
    \label{generalpropagator}
\end{equation}
where each of the coefficients have the structure
\begin{equation}
    D_i=\frac{G}{1+G P_i},
    \label{generalcoefficients}
\end{equation}
with $i=\parallel, \ \perp$ and $0$ and $G=(p_0^2-p_\perp^2-p_3^2-M_\rho^2)^{-1}$. We remind you that $M_\rho$ is the mass of the neutral rho-meson. Therefore, the explicit form for propagator is
\begin{align}
    D_\rho^{\mu\nu}&=\frac{\Pi_\parallel^{\mu\nu}}{p_0^2-p_\perp^2-p_3^2-M_\rho^2-P_\parallel}\nonumber \\
    &+\frac{\Pi_\perp^{\mu\nu}}{p_0^2-p_\perp^2-p_3^2-M_\rho^2-P_\perp}\nonumber \\
    &+\frac{\Pi_0^{\mu\nu}}{p_0^2-p_\perp^2-p_3^2-M_\rho^2-P_0}.
    \label{propagatorexplicit}
\end{align}
From Eq.~(\ref{propagatorexplicit}), we notice that there are three different modes: parallel, perpendicular and zero. Each of them is a mode for the screening mass, which can be computed from the general equations
\begin{equation}
    p_0^2-p_\perp^2-p_3^2-M_\rho^2-\Re[P_i(p_0,p_\perp,p_\parallel)]=0,
    \label{scateringrelation}
\end{equation}
\begin{figure}[t]
    \centering
    \includegraphics[scale=0.55]{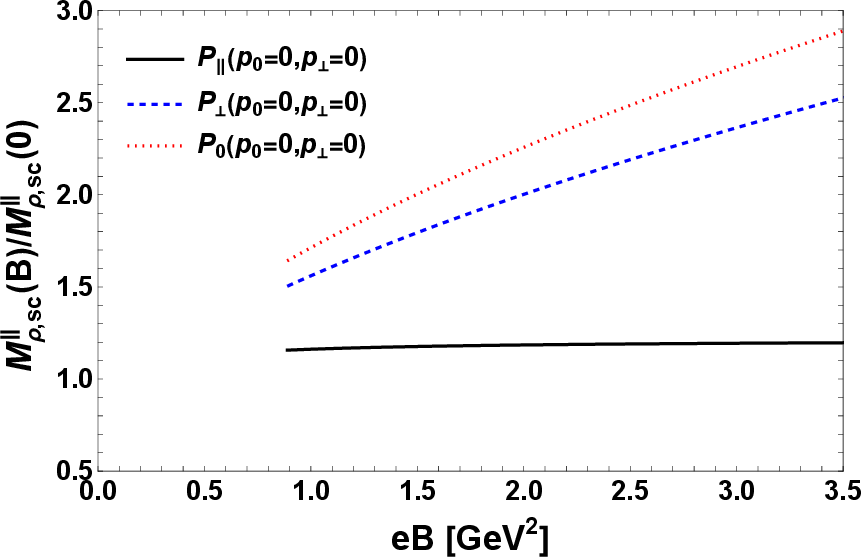}
    \caption{The three modes of the parallel screening mass of the neutral rho-meson, normalized to the rho-meson mass in vacuum, are shown as a function of the magnetic field strength. The red dotted curve corresponds to the zero mode, the blue dashed curve represents the perpendicular mode, and the solid black curve depicts the behavior of the parallel mode.}
    \label{screeningmassparallel}
\end{figure}
when $p_0 \rightarrow 0$, and for $i=\parallel, \ \perp$ and $0$. Once $p_0$ has become zero, we see that we have two cases, the parallel screening mass (with $p_\perp \rightarrow 0$ and $p_3 \neq 0$) and the perpendicular screening mass (with $p_3 \rightarrow 0$ and $p_\perp \neq 0$). For the parallel screening mass the equations to solve are
\begin{align}
    p_3^2&=-M_\rho^2-\Re[P_\parallel(p_0=0,p_\perp=0,p_3)],\nonumber \\
    p_3^2&=-M_\rho^2-\Re[P_\perp(p_0=0,p_\perp=0,p_3)], \nonumber \\
    p_3^2&=-M_\rho^2-\Re[P_0(p_0=0,p_\perp=0,p_3)],
    \label{parallelequation}
\end{align}
for the unknown $p_3$, where $(M_{\rho,sc}^{\parallel})^2\equiv-p_3^2$, and for the perpendicular screening mass, we have
\begin{align}
    p_\perp^2&=-M_\rho^2-\Re[P_\parallel(p_0=0,p_\perp,p_3=0)], \nonumber \\
    p_\perp^2&=-M_\rho^2-\Re[P_\perp(p_0=0,p_\perp,p_3=0)], \nonumber \\
    p_\perp^2&=-M_\rho^2-\Re[P_0(p_0=0,p_\perp,p_3=0)],
    \label{perpequation}
\end{align}
with the unknown $p_\perp$, where $(M_{\rho,sc}^{\perp})^2\equiv-p_\perp^2$, where $P_0(p_0,p_\perp,p_3)$, $P_\perp(p_0,p_\perp,p_3)$ and $P_\parallel(p_0,p_\perp,p_3)$ correspond to Eqs.~(\ref{finalP0}), (\ref{finalPperp}) and~(\ref{finalPparallel}), respectively.
The solutions of Eq.~(\ref{parallelequation}) are depicted in Fig.~(\ref{screeningmassparallel}), where we plot the three modes of the parallel screening mass as a function of the magnetic field strength. Similarly, the solutions of Eq.~(\ref{perpequation}) are shown in Fig.~(\ref{screeningmassperp}), where we also plot the three modes of the perpendicular screening mass as a function of the magnetic field strength.

\begin{figure}[t]
    \centering
    \includegraphics[scale=0.55]{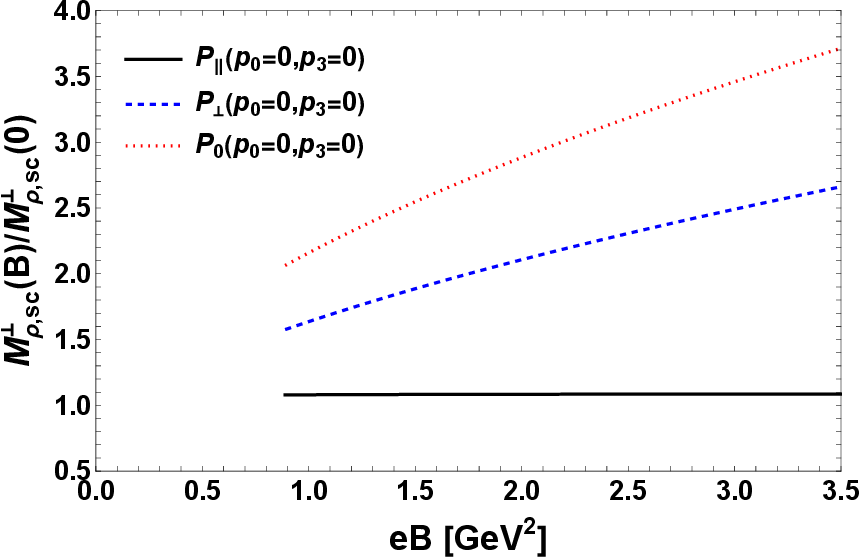}
    \caption{The three modes of the perpendicular screening mass of the neutral rho-meson, normalized to the rho-meson mass in vacuum, are shown as a function of the magnetic field strength. The red dotted curve corresponds to the zero mode, the blue dashed curve represents the perpendicular mode, and the solid black curve depicts the behavior of the parallel mode.}
    \label{screeningmassperp}
\end{figure}

Figure~(\ref{screeningmassparallel}) shows the three modes of parallel screening mass of the neutral rho-meson, normalized to its value at a zero magnetic field, as a function of the magnetic field intensity. We use the following parameter values for the model: $M_\rho=770$ MeV, $m_\pi=140$ MeV, and $g^2/4\pi=2.93$~\cite{Gale:1990pn}. Additionally, we set $\mu=2eB$ to satisfy the lowest Landau level approximation. As a function of the magnetic field, we observe a monotonically increasing behavior for the zero and perpendicular modes, whereas in the parallel mode, the increase is very small. For the perpendicular screening mass of the neutral rho-meson, the three modes are depicted in Fig.~(\ref{screeningmassperp}), using the same parameter values. Once again, the zero and perpendicular modes exhibit a monotonically increasing behavior with the magnetic field, while in the parallel mode, the increase remains minimal.

\section{\label{sec5} Conclusions}

In this work, we have analyzed the behavior of the screening mass of the rho-meson as a function of the magnetic field strength. We computed the neutral rho-meson self-energy at one-loop order using the lowest Landau level approximation, assuming that the field strength is the dominant energy scale in the physical system. The model employed is the KLZ model, in which the degrees of freedom are the charged pions and the neutral rho-mesons. Since the neutral rho-meson is a vector field, both its propagator and self-energy are tensors. However, because this model is gauge-invariant, both quantities must be transverse.

Moreover, in this study, we included a uniform and constant magnetic field, which induces Lorentz symmetry breaking. As a consequence, we expressed both the propagator and the self-energy in terms of three linearly independent tensor structures, $\Pi^{\mu\nu}_\parallel$, $\Pi^{\mu\nu}_\perp$ and $\Pi^{\mu\nu}_0$, leading to three distinct modes for the screening mass. Lorentz symmetry breaking also causes a separation of the four-momentum components into parallel and perpendicular directions, resulting in two types of screening masses for the neutral rho-meson: the parallel screening mass ($p_0=0$ and $p_\perp\rightarrow0$) and the perpendicular screening mass ($p_0=0$ and $p_\parallel\rightarrow0$). We find that the zero and perpendicular modes for both screening masses exhibit an increasing behavior as a function of the magnetic field strength, whereas in the parallel mode, the increase is very small.. The concept of screening mass originates from linear response theory, where it describes how a medium reacts to an external static field. Due to the static nature of the field, its screening within the medium is governed by the system’s response function in the limit where $p_0=0$ and $\vec{p}$ remains finite. The inverse of the screening mass corresponds to the screening or Debye length, which characterizes how the medium attenuates the external perturbation. Consequently, we find that, in all cases studied, the Debye length decreases as the magnetic field increases. Finally, another important conclusion from this work arises from the well-established relation between the pole mass and the corresponding screening masses when $eB\neq 0$~\cite{Sheng:2020hge}. In particular, it holds that $M_{\rho,pole}=M_{\rho,sc}^\parallel$ and $M_{\rho,sc}^\parallel\neq M_{\rho,sc}^\perp$. From Ref.~\cite{Bali:2017ian}, we observe that the pole mass of the neutral rho-meson exhibits an increasing behavior as the magnetic field increases, regardless of the spin projection. Therefore, our results are qualitatively consistent with those found for the pole mass of the neutral rho-meson.

\begin{acknowledgments}

Support for this work was received in part by the Consejo Nacional de Humanidades, Ciencia y Tecnología Grant No. CF-2023-G-433. LAH acknowledges support from the DCBI UAM-I PEAPDI 2024, and DAI UAM PIPAIR 2024 projects under Grant No. TR2024-800-00744. R.Z acknowledges support from ANID/CONICYT FONDECYT Regular (Chile) under Grant No. 1241436, and DM-S acknowledges the financial support of a fellowship granted by Consejo Nacional de Humanidades, Ciencia y Tecnología as part of the Sistema Nacional de Posgrados. 

\end{acknowledgments}




\nocite{*}

\bibliography{mybibliography}

\end{document}